# Quantitative Impact of Occupation Smearing Methods on Conventional Superconductivity: from Electronic Structure through Lattice Dynamics to Electron-Phonon Coupling

Zhongyu Wan

*Department of Physics, City University of Hong Kong, Hong Kong SAR, 999077, People's Republic of China*

*Corresponding email:zywan5-c@my.cityu.edu.hk*

**Abstract**

In first-principles calculations of metallic systems, the discrete electronic mesh in the Brillouin zone necessitates a smearing function and its width (e.g., the widely used 0.02 Ry Methfessel-Paxton scheme) to enforce continuity of energy and occupation at the Fermi surface. However, the quantitative impact of such empirical parameters on phonon-mediated superconductivity has long been overlooked, leading to their habitual selection by mere convention. Here, using hexahydride $CsH_6$ as a benchmark, we systematically investigate the magnitude and propagation mechanism of smearing-induced errors against the parameter-free optimized tetrahedron method. Long-tailed smearing delocalizes the occupation window, causing substantial Fermi-level shifts, flattening of the density of states, and hardening phonon modes, thereby weakening electron-phonon coupling. Conversely, short-tailed smearing suffers from insufficient sampling, producing density-of-states oscillations and enhanced Fermi-surface nesting; via amplified electronic screening and Peierls instabilities, this softens acoustic modes and generates imaginary frequencies, causing $CsH_6$ to be dynamically unstable and discarded. However, the optimized tetrahedron method reveals that $CsH_6$ is dynamically stable under ambient pressure with a remarkable superconducting transition temperature of 121 K. To prevent subjective empirical parameters from obscuring theoretically promising high-temperature superconductors, we derive a simple scaling relation for the smearing width, applicable to different smearing functions, from the free-electron-gas model, offering an objective, precise, and efficient protocol for parameter selection in high-throughput screening.

## 1. **Introduction**

Traditionally, new materials discovery via trial-and-error approaches has relied on empirical chemical intuition and protracted synthesis–characterization cycles, substantially elevating research and development costs[1]. With advances in computational techniques and theoretical methodologies, first-principles calculations based on density functional theory have emerged as a leading paradigm in materials design[2-4]. By solving the Kohn–Sham equations at the atomic scale[5], these methods enable predictive modeling of critical properties—including electrical and thermal conductivity[6], optical response[7], and electron–phonon coupling[8-10] (EPC)—thereby accelerating the screening and design of advanced materials, all of which are fundamentally dictated by their underlying electronic structure[11].

Band structure classifies solids as metals or insulators based on the energy gap between the valence band maximum and conduction band minimum evident in the density of states (DOS)[12]. For most insulators, the band gap is the defining characteristic, so the precise form of electronic occupation and the integration scheme over the Brillouin zone are of minor importance[13]. In metallic systems, however, discrete k-point sampling yields bands that cross the Fermi level ($\varepsilon_F$), producing an abrupt 1→0 transition in electron occupation near the Fermi surface[14]. Numerical occupation methods are therefore required to enforce a continuous energy distribution and to handle this discontinuity in a manner consistent with Fermi-Dirac statistics[15]. The optimized tetrahedron method is currently the only parameter-free approach capable of accurately converging integrals of response functions containing Fermi-surface singularities[16,17]. It has provided the most reliable descriptions for strongly electron-phonon-coupled superconductors such as classic $MgB_2$ and high-pressure fcc Li,[17] and has accordingly been adopted as the absolute benchmark in numerous validation studies[18,19]. Nevertheless, this method becomes prohibitively expensive for high-throughput calculations involving thousands of structures[20].

Currently, the prevailing approach is to regularize the physical Fermi-Dirac statistics with a simple finite-width approximation (i.e., a smearing function) in place of discontinuous occupation[14,21]. By tuning the smearing width $\sigma$, this technique smooths the Fermi surface over discrete $k$-point grids and stabilizes self-consistent field convergence[22]. While this artificial parameter dramatically reduces computational costs, it introduces two inherent empirical drawbacks: First, different formulations of smearing functions prioritize distinct physical aspects, leading to systematic discrepancies in the fine profiles and weight allocations of electronic states[23]. Second, the choice of smearing width determines the energy window over which electrons at each $k$-point are broadened[18]. Consequently, approximating the smearing function inevitably alters the subtle details of electron distribution near the Fermi surface (Figs. 1(a)~(d)). This artifact directly feeds into the estimation of electron-phonon coupling and can propagate to lattice dynamics via the Peierls mechanism. As a result, promising high $T_c$ superconductor candidates risk being erroneously screened out during the early stages of high-throughput calculations[24], ultimately leading to an underestimation of the dataset's overall superconducting potential.

Regrettably, this critical issue has been largely overlooked in numerous computational workflows[25-27]. The absence of systematic understanding and benchmarking regarding how phonon-mediated properties evolve with empirical parameters, compounded by past discussions limited merely to the convergence of total energy and forces[23], renders the selection of smearing functions and widths in current density functional calculations subjective or even arbitrary[24,28-31]. Theoretically predicted high-temperature superconductors are particularly promising candidates, yet they are typically characterized by pronounced van Hove singularities (VHS) densely distributed near the Fermi level[32,33]. Consequently, numerical artifacts arising from computational approximations can severely compromise the predictive reliability for such superconductors. These concerns naturally raise three fundamental questions: (1) What are the quantitative discrepancies between mainstream smearing methods and the optimized tetrahedron method in describing electronic structure, phonon dispersion, and superconducting properties? (2) What is the quantitative mechanism by which smearing-induced errors propagate through these hierarchical properties? (3) What objective criterion should govern the selection of smearing parameters to eliminate subjectivity in predictive results?

To address these questions, we identify the recently proposed cubic hexahydride as an ideal platform[9,10]. Its promising ambient-pressure high-temperature superconductivity originates from strong electron-phonon coupling, making it particularly susceptible to amplifying subtle numerical deviations into pronounced physical variations. This sensitivity offers a unique perspective for quantifying the scaling relationship between occupation approximations and phonon-mediated superconductivity.

## 2. Theory

Since the calculation of EPC is formulated within the linear-response framework[34], the eigenvalues of the phonon dynamical matrix are first obtained via density-functional perturbation theory[35]:

$$\omega^2 \propto D = D_{ion} + D_{ele} \tag{1}$$

Where $\omega$ is the phonon frequency, $D_{ion}$ is the direct Coulomb repulsion between atomic nuclei, which tends to produce high frequencies, $D_{ele}$ is the shielding contribution generated by the electron cloud, and tends to soften phonon frequencies[36,37]. Furthermore, for the $D_{ele}$ has:

$$D_{ele} \propto \frac{1}{M}\sum_{\mathbf{k},n,m}\frac{f_{n\mathbf{k}}-f_{m\mathbf{k+q}}}{\varepsilon_{n\mathbf{k}}-\varepsilon_{m\mathbf{k+q}}}\left|\langle\psi_{n\mathbf{k}}\,|\frac{\partial V_{SCF}}{\partial \mathbf{u_q}}|\,\psi_{m\mathbf{k+q}}\rangle\right|^2 \tag{2}$$

Where $(f_{n\mathbf{k}}-f_{m\mathbf{k+q}})$ is the difference in the number of electrons occupied by the *n*-th and *m*-th bands at the electron wavevector **k** under phonon vector **q** perturbations, and $\varepsilon$ is the eigenenergy of the electron. For the same Fermi surface, $m = n$, when $\mathbf{q}\rightarrow 0$ and $\varepsilon \rightarrow \varepsilon_F$,

$$\lim_{\mathbf{q}\rightarrow 0}\frac{f_{n\mathbf{k}}-f_{m\mathbf{k+q}}}{\varepsilon_{n\mathbf{k}}-\varepsilon_{m\mathbf{k+q}}} \approx \left(\frac{\partial f}{\partial \varepsilon}\right)_{\varepsilon_F} \tag{3}$$

Where $\frac{\partial f}{\partial \varepsilon}$ is the smearing function *g* at $\varepsilon_F$, which demonstrates that the form of electronic occupation affects phonon frequencies even when the k-point mesh is sufficiently dense. For the Gaussian smearing approximation at energy $\varepsilon$, the occupation function *f* and the smearing function *g* are given by[38]:

$$f = \frac{1}{2}\left[1-\mathrm{erf}\left(\frac{\varepsilon-\varepsilon_F}{\sigma}\right)\right] = \frac{1}{2}\mathrm{erfc}\left(\frac{\varepsilon-\varepsilon_F}{\sigma}\right) \tag{4}$$

$$g = \frac{1}{\sqrt{\pi}}e^{-(\varepsilon-\varepsilon_F)^2/\sigma^2} \tag{5}$$

The occupation function and smearing function for the first-order Methfessel-Paxton (MP) smearing are respectively given by[39]:

$$f = \frac{1}{2}\mathrm{erfc}\left(\frac{\varepsilon-\varepsilon_F}{\sigma}\right) - \frac{1}{4\sqrt{\pi}}\left(\frac{\varepsilon-\varepsilon_F}{\sigma}\right)\exp\left[-\left(\frac{\varepsilon-\varepsilon_F}{\sigma}\right)^2\right] \tag{6}$$

$$g = \frac{1}{\sqrt{\pi}\sigma}e^{-\left(\frac{\varepsilon-\varepsilon_F}{\sigma}\right)^2}\left[1+\frac{1}{4}(4\left(\frac{\varepsilon-\varepsilon_F}{\sigma}\right)^2-2)\right] \tag{7}$$

The occupation function and smearing function for the Marzari-Vanderbilt (MV) cold smearing method take the following forms, respectively[40]:

$$f = \frac{1}{2}\mathrm{erfc}\left(\left(\frac{\varepsilon-\varepsilon_F}{\sigma}\right)-\frac{1}{\sqrt{2}}\right) + \frac{1}{\sqrt{2\pi}}\exp\left[-\left(\left(\frac{\varepsilon-\varepsilon_F}{\sigma}\right)-\frac{1}{\sqrt{2}}\right)^2\right] \tag{8}$$

$$g = \frac{1}{\sqrt{\pi}}\left(2-\sqrt{2}\left(\frac{\varepsilon-\varepsilon_F}{\sigma}\right)\right)\exp\left[-\left(\left(\frac{\varepsilon-\varepsilon_F}{\sigma}\right)-\frac{1}{\sqrt{2}}\right)^2\right] \tag{9}$$

For Fermi-Dirac (FD) smearing, the occupation function and smearing function are given by[41]:

$$f = \frac{1}{e^{(\varepsilon-\varepsilon_F)/\sigma}+1} \tag{10}$$

$$g = \frac{1}{\sigma}\frac{e^{(\varepsilon-\varepsilon_F)/\sigma}}{\left[e^{(\varepsilon-\varepsilon_F)/\sigma}+1\right]^2} = \frac{1}{4\sigma}\operatorname{sech}^2\left(\frac{\varepsilon-\varepsilon_F}{2\sigma}\right) \tag{11}$$

Here, $\sigma$ corresponds to the physical electron temperature in the FD smearing scheme, namely $\sigma = k_B T$, whereas in the other three smearing methods it serves merely as a numerical parameter. By introducing the dimensionless variable $x=(\varepsilon-\varepsilon_F)/\sigma$, the distributions of $g$ and $f$ relative to the Fermi level can be visualized (Figs. 1(e), (f)). Furthermore, the electronic density of states at energy $\varepsilon$ is evaluated as[42]:

$$D(\varepsilon) = \sum_{n,\mathbf{k}} w_{\mathbf{k}} \cdot \frac{1}{\sigma} g\left(\frac{\varepsilon-\varepsilon_{n\mathbf{k}}}{\sigma}\right) \tag{12}$$

where $w_{\mathbf{k}}$ denotes the weight of wave vector $\mathbf{k}$ in the Brillouin zone. It follows that the electronic population at energy $\varepsilon$ depends on both the choice of smearing function and its width. In the limit of sufficiently dense k-point sampling where convergence is achieved, Eq. (12) reduces from a discrete summation to an ideal integral form:

$$D(\varepsilon) = \sum_{n} \frac{V_{\text{cell}}}{(2\pi)^3}\int_{BZ} d\mathbf{k}\, \frac{1}{\sigma} g\left(\frac{\varepsilon-\varepsilon_{nk}}{\sigma}\right) \tag{13}$$

where $V_{\text{cell}}$ is the unit cell volume. In the limit of sufficiently dense k-point sampling, the density of states becomes an exact integral over the Brillouin zone; it is theoretically independent of the k-point mesh density and determined solely by the smearing function. Together with the charge conservation condition[36]:

$$N_{\text{tot}} = \sum_{n,\mathbf{k}} w_{\mathbf{k}} f\left(\frac{\varepsilon-\varepsilon_F}{\sigma}\right) \tag{14}$$

where $N_{\text{tot}}$ represents the total electron count of the compound. For a sufficiently dense k-point mesh, the discrete sum converges to the integral:

$$N_{\text{tot}} = \int_{-\infty}^{\varepsilon_F} N(\varepsilon)d\varepsilon = \sum_{n,\mathbf{k}} w_{\mathbf{k}} \int_{-\infty}^{\varepsilon_F} \frac{1}{\sigma} g\left(\frac{\varepsilon-\varepsilon_{n\mathbf{k}}}{\sigma}\right) d\varepsilon \tag{15}$$

With the $N_{\text{tot}}$ held constant, the Fermi level $\varepsilon_F$ is found to depend explicitly on both the smearing width $\sigma$ and the functional form $g$. Further computational details are provided in the Supporting

Information.

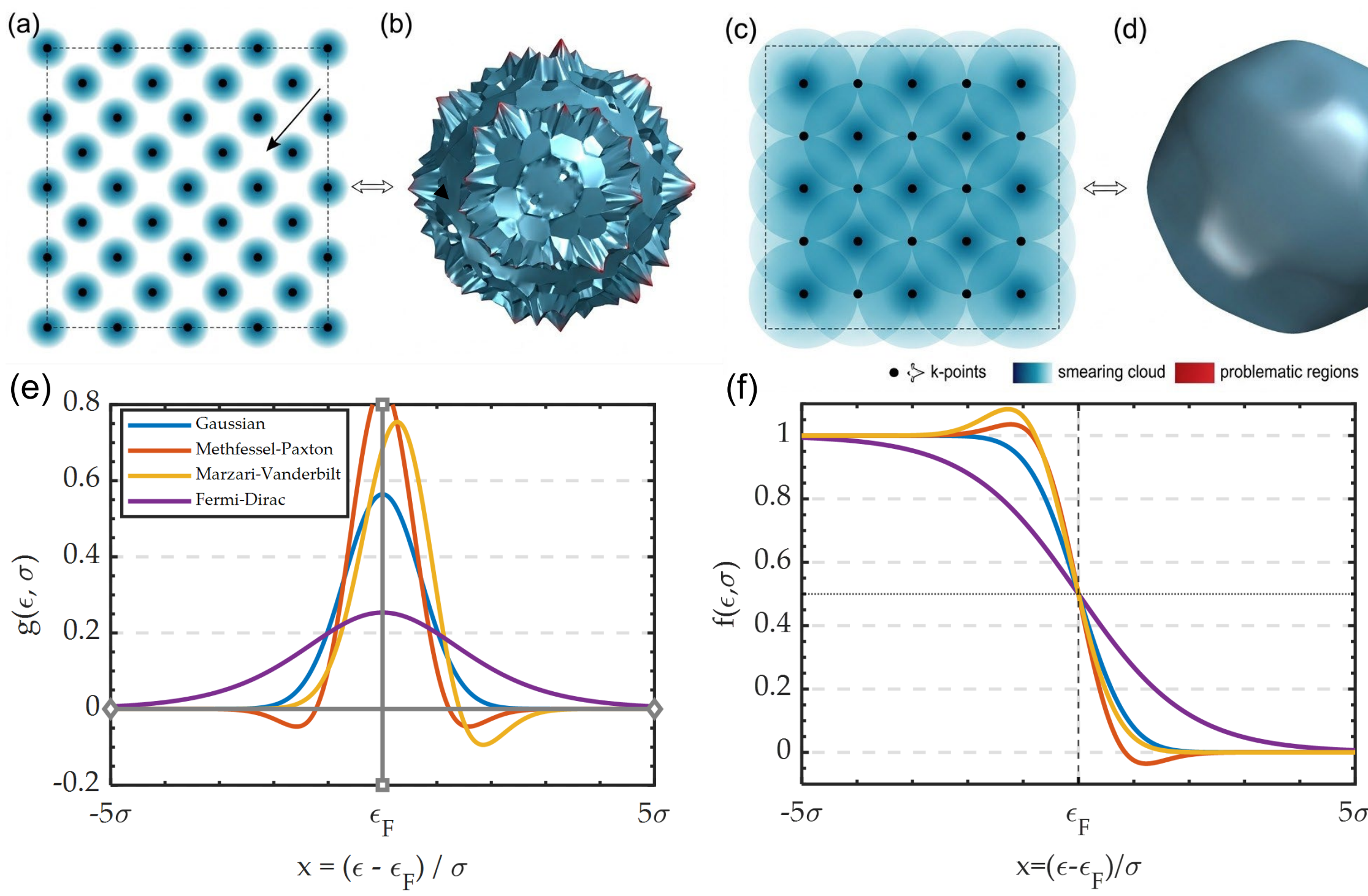

**Figure 1.** Schematic illustrations of (a, b) Fermi surface undersampling due to the smearing short-tail effect and (c, d) Fermi surface over-smoothing due to the long-tail effect. Comparison of energy windows for different (e) smearing functions and (f) occupancy functions.

## 3. Results and Discussion

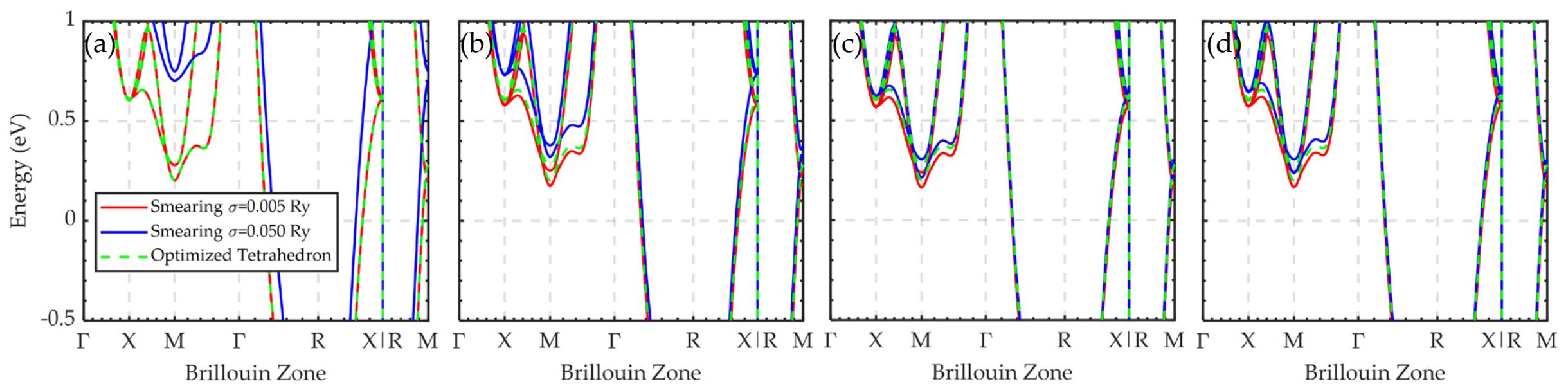

**Figure 2.** Shifts in the electronic band structure induced by smearing using (a) Fermi-Dirac, (b) Gaussian, (c) Methfessel–Paxton, and (d) Marzari–Vanderbilt smearing functions.

Considering a 95% effective occupancy, the widths of the FD, Gaussian, MP, and MV functions are $7.33\sigma$, $2.77\sigma$, $1.49\sigma$, and $1.91\sigma$, respectively (Figure 1(e)). Consequently, both the functional form and the width of the smearing function dictate the energy window over which electrons are smeared out. A small $\sigma$ or the use of the MP function yields a short-tailed smearing effect, whereas a large $\sigma$ or the implementation of the FD function leads to a long-tailed smearing behavior.

Equation (14) clearly demonstrates that both the functional form and the width of the smearing scheme shift the relative position of the $\varepsilon_F$ in the electronic structure. For a minimal width of $\sigma = 0.005$ Ry, the FD approximation yields the most ideal alignment (Figure 2). However, as $\sigma$ increases to $0.050$ Ry, the entire band structure is significantly shifted upward away from $\varepsilon_F$. Conceptually, this occurs because the FD distribution intrinsically possesses a long-tailed nature; a small $\sigma$ is thus necessary to offset the over-dispersion of electronic occupancy under the FD approximation. Notably, the effective width of the FD function is $7.33\sigma$—substantially larger than the scaling factors of the other three functions. This disproportionately large effective width accounts for the acute sensitivity of $\varepsilon_F$ to $\sigma$ under FD smearing, while also explaining the relative robustness of $\varepsilon_F$ observed in Gaussian, MP, and MV schemes.

Nevertheless, regardless of the smearing function employed, an increased $\sigma$ consistently drives flat bands away from $\varepsilon_F$—a trend also observed in $MgB_2$ and fcc Li[17]. Physically, this behavior arises because the long-tailed effect encompasses states across a broader energy window, thereby populating higher energy levels. To maintain charge neutrality, the system's $\varepsilon_F$ must undergo a pronounced shift toward regions of lower band density[43].

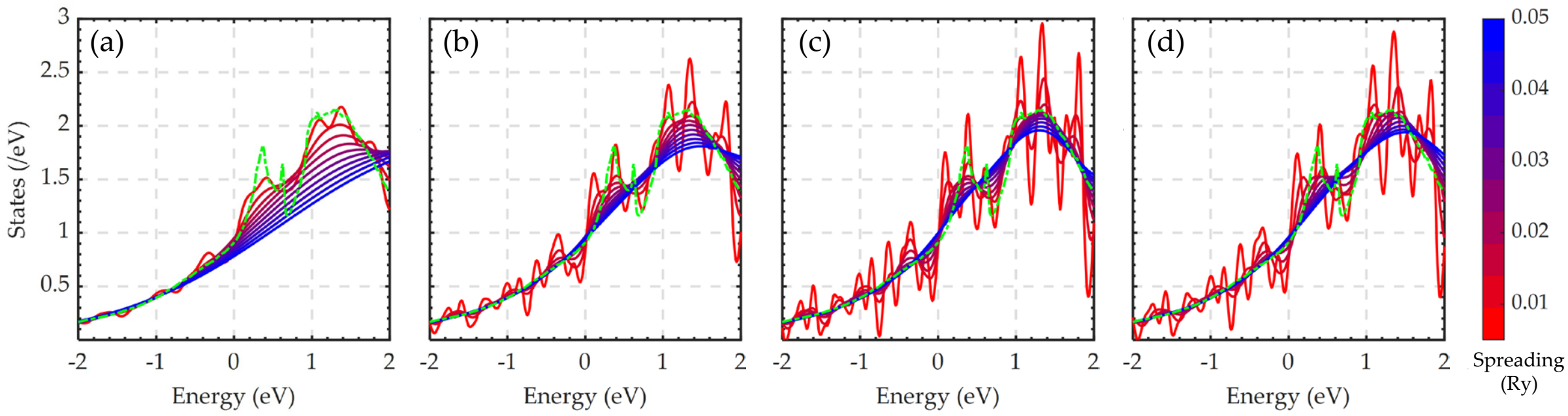

**Figure 3.** Evolution of the electronic density of states with respect to smearing width for different smearing schemes: (a) Fermi–Dirac, (b) Gaussian, (c) Methfessel–Paxton, and (d) Marzari–Vanderbilt. The optimized tetrahedron method, serving as the benchmark, is shown in green.

Equation (12) indicates another direct consequence of differing smearing approximations: pronounced discrepancies in the precision of the DOS. At a fixed width $\sigma$, the FD scheme yields the smoothest DOS owing to its long-tailed distribution, which blurs the finer features of VHS (Figure 3). Conversely, the narrow effective widths intrinsic to the MP and MV formulations, combined with small artificial $\sigma$ values, induce severe DOS oscillations and introduce spurious VHS artifacts. Physically, the short-tailed effect localizes the weight distribution of occupied states at $\varepsilon_{n\mathbf{k}}$, driving the wavefunctions across the discrete $k$-point grid from a continuous distribution toward a localized representation and thereby triggering divergence in the state integration (Figure 1(a)). Consequently, applying larger $\sigma$ values within the MP and MV schemes significantly alleviates these DOS oscillations. This improvement occurs because electrons at $\varepsilon_{n\mathbf{k}}$ are dispersed over a broader energy range, mitigating the $k$-point grid undersampling issue (Figure 1(c)).

Equations (1) and (2) reveal the explicit dependence of phonon frequencies on electronic occupations, directly highlighting the impact of occupancy approximations on lattice dynamics. Decreasing $\sigma$ sharpens the derivative of the occupation function at the Fermi level, $\partial f/\partial \varepsilon$ (Figure 1(f)), which enhances electronic screening near $\varepsilon_F$ (Equations (2) and (3)). This, in turn, distorts the interatomic force constants and induces phonon softening (Equation (1)). Conceptually, the sharp DOS fluctuations near $\varepsilon_F$ caused by the short-tailed effect correspond to a steep slope in the variation of electronic states (Figure 3). Within the framework of perturbation theory, abrupt shifts in electronic state counts increase the matrix elements of the electronic potential derivative[44]. This manifests as a pronounced enhancement in the EPC strength ($\lambda$), thereby exacerbating structural instabilities in metallic systems[45]. This mechanism also naturally accounts for the systematic trend reported in existing benchmarks[17], where $\lambda$ increases monotonically as $\sigma$ decreases.

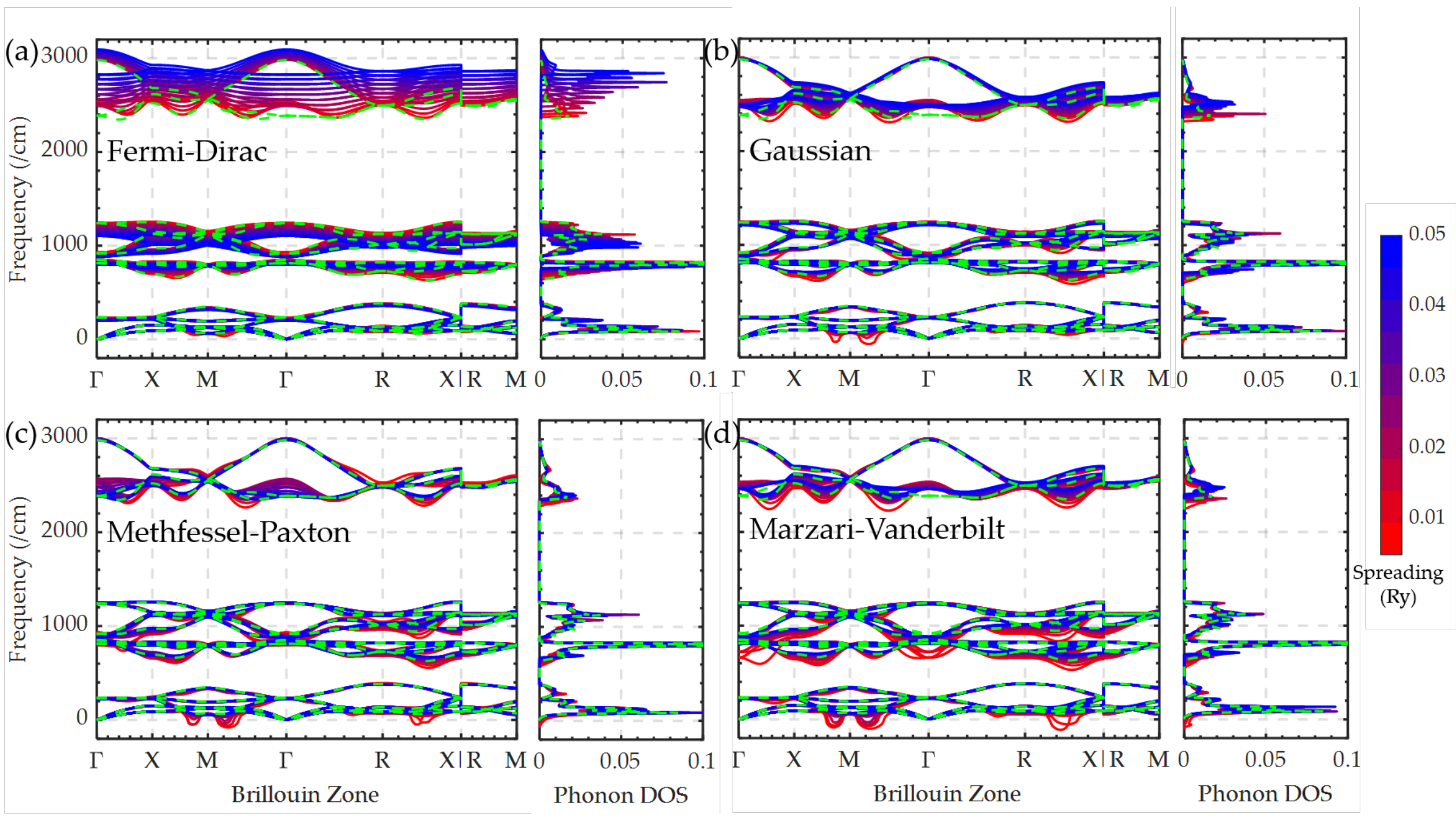

**Figure 4.** Variation of phonon dispersions with respect to smearing width under different smearing functions: (a) Fermi–Dirac, (b) Gaussian, (c) Methfessel–Paxton, and (d) Marzari–Vanderbilt. The benchmark optimized tetrahedron method is depicted by green dashed lines.

As depicted in the phonon dispersions (Figure 4), at a minimal width of $\sigma = 0.005$ Ry, both the MP and MV schemes induce imaginary frequencies in the acoustic branches along the $\mathrm{X-M-\Gamma}$ and $\mathrm{R-X}$ high-symmetry paths. The Gaussian scheme exhibits unstable modes only along $\mathrm{X-M-\Gamma}$, whereas the FD approximation maintains continuous dynamical stability. As $\sigma$ increases, the imaginary frequency modes across all smearing functions are eliminated. This stabilization occurs because the DOS flattening induced by a larger $\sigma$ weakens the electronic screening effect (Equation (2) and Figure(3)) and drives electronic states along $\mathrm{X-M-\Gamma}$ away from $\varepsilon_F$ (Figure (2)). This hardening mechanism of the acoustic modes is inherently attributable to the long-tailed effect of the occupation approximation.

However, varying $\sigma$ under the FD approximation exerts a negligible effect on the acoustic modes (Figure 4 (a)); instead, notable perturbations manifest in the intermediate-frequency (600–1300 $cm^{-1}$) and high-frequency (2300–3200 $cm^{-1}$) optical branches. Because $\sigma$ in the FD approximation physically corresponds to the electronic temperature[41], the lighter H atoms undergo more pronounced thermal excitations than the heavier Cs atoms, confining the predominant frequency shifts to the H-dominated optical modes. In contrast, the optical branches under the Gaussian and MP approximations remain relatively robust against variations in $\sigma$ (Figure 4 (b) (c)).

Notably, the entire phonon dispersion under the MV scheme undergoes systematic hardening as $\sigma$ increases (Figure 4 (d)). Given that the occupation function $f(x)$ of the MV approximation consistently exceeds those of other schemes at energy levels below $x < -1\sigma$ (Figure 1(f)), the MV smearing inherently introduces additional occupations[40]. As $\sigma$ expands, these extra occupied states push deeper into lower energy levels, effectively mimicking system compression and hardening all chemical bonds.

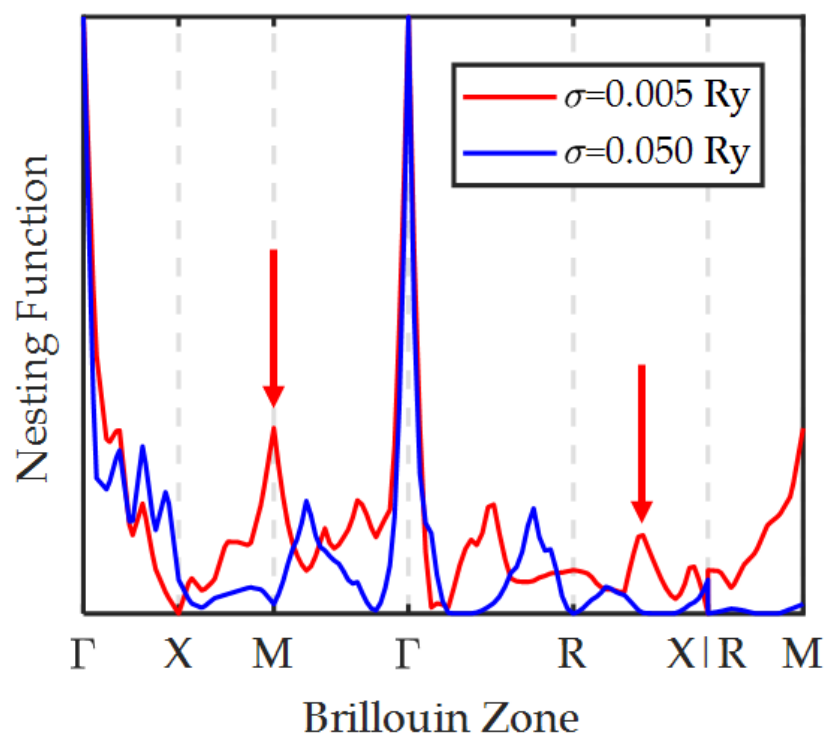

**Figure 5.** Smearing-width-induced variations of the Fermi surface nesting function, where the double-delta function is approximated using the Methfessel–Paxton formula.

Apart from the influence of DOS modifications induced by smearing approximations, discrepancies in electronic occupations across the Fermi surface are directly reflected in the nesting function, which further modulates dynamical stability via the Peierls distortion mechanism[45]. Given that the double-delta function in density functional perturbation theory calculations is likewise treated using smearing schemes[44], we take the widely used MP function as a representative case (Figure 5). At a minimal width of $\sigma = 0.005$ Ry, the nesting function along the $\mathrm{X-M-\Gamma}$, $\mathrm{R-X}$, and $\mathrm{R-M}$ paths is significantly enhanced compared to that at $\sigma = 0.05$ Ry. Furthermore, prominent nesting peaks emerge at the M point and near the midpoint of the $\mathrm{R-X}$ line. This behavior indicates that a short-tailed smearing width strengthens Fermi surface nesting at these specific **k**-vectors, ultimately softening the acoustic phonon modes into imaginary frequencies. Conversely, a long-tailed width attenuates the Peierls instability driven by Fermi surface nesting.

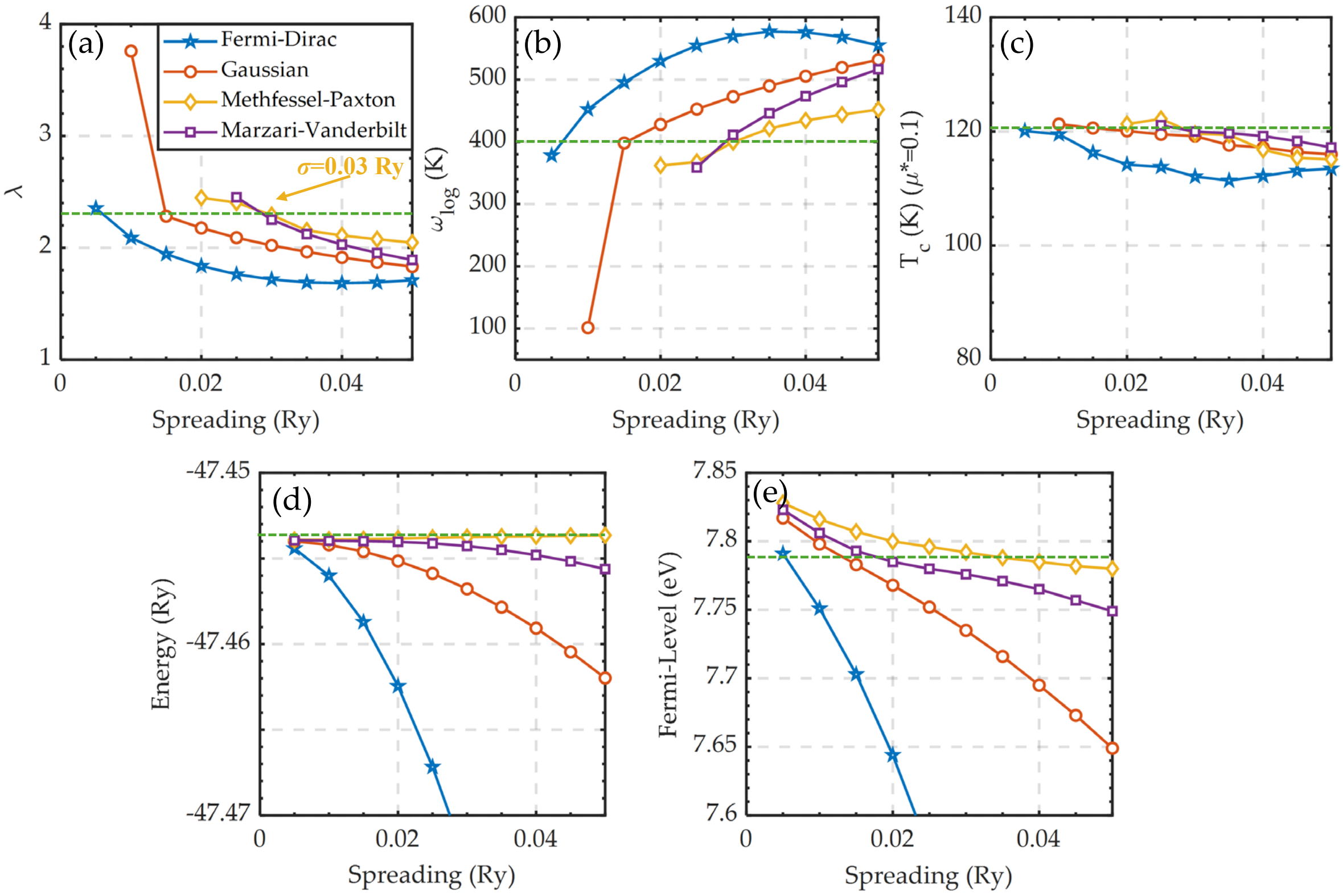

**Figure 6.** Smearing-width dependence of EPC properties across different smearing schemes: (a) EPC strength, (b) logarithmic average phonon frequency, (c) isotropic Migdal–Eliashberg superconducting transition temperature ($\mu$*=0.1), (d) total energy, and (e) Fermi energy. The benchmark optimized tetrahedron method is highlighted by green dashed lines.

Fermi surface nesting further propagates into the predictions of phonon-mediated superconductivity. Under the premise of dynamical stability, the FD approximation systematically yields the lowest $\lambda$ and the highest logarithmic average phonon frequency ($\omega_{\log}$), followed by the Gaussian smearing (Figure 6 (a) (b)). This trend directly stems from their respective degrees of electron smearing. Evidently, the calculated $\lambda$ across all methods exhibits a monotonic decrease with increasing $\sigma$, consistent with trends reported in existing benchmarks[17,28-30]. This behavior results from the synergistic interplay of band structure effects—namely, the shift of flat bands away from $\varepsilon_F$ and the flattening of the DOS—together with the systematic hardening of the phonon spectrum (Figure (2)~(4)).

Furthermore, within the range of $\sigma = 0.025–0.05$ Ry (Figure 6 (a) (b)), the variations in $\lambda$ and $\omega_{\log}$ calculated via the Marzari-Vanderbilt scheme are more pronounced than those obtained from the Methfessel-Paxton. This difference arises because an increased $\sigma$ induces a global blueshift across the entire phonon spectrum in the MV approximation, whereas phonon hardening in the MP scheme remains largely confined to the acoustic branches (Figure (4)).

Nevertheless, the estimation of $T_c$ for $CsH_6$ appears remarkably satisfactory: the deviations across the four smearing schemes range from $-9.4$ to $+1.8$ K ($-7.8\%$ to $+1.2\%$), showing excellent agreement with the benchmark value of 120.8 K obtained via the optimized tetrahedron method (Figure 6 (c)). Although both the choice of smearing function and its width significantly alter the calculated $\lambda$, the concurrent evolution of $\omega_{\log}$ acts as a competing factor that counterbalances the impact of $\lambda$ on $T_c$—a cancellation mechanism also observed in the prototypical $H_3S$ system[46].

Consequently, we propose that the robustness of $T_c$ is intimately tied to the local features of the DOS at $\varepsilon_F$. For instance, $T_c$ in $CaH_6$ exhibits negligible variations across $\sigma = 0.005$-$0.05$ Ry because its $\varepsilon_F$ resides in a DOS valley[24]. In contrast, for $Mg_2IrH_6$ at a comparable $k$-point density[47], $\varepsilon_F$ perfectly aligns with a VHS, resulting in pronounced $T_c$ fluctuations that require larger $\sigma$ values or denser $k$-point grids to reach a convergence plateau[24]. This high sensitivity stems from the zero band gradient and divergent DOS at the VHS, which necessitate refined $k$-point sampling and broader smearing to smooth out the singularity. In our present case, $CsH_6$ lies merely along the slope of a DOS peak (Figure 3), explaining its mild $T_c$ fluctuations.

Furthermore, we observe that the FD function systematically yields the lowest $T_c$ values with the largest variations (Figure 6 (c)). It is well established that $T_c$ is strongly correlated with the electron density at $\varepsilon_F$[48]. Under FD smearing, this suppression arises from two factors: the upward shift of electronic bands along the $\mathrm{X}-\mathrm{M}-\Gamma$ path (Figure 2 (a)), and the broad energy window that scatters electrons near the Fermi surface into distant orbital states (Figure 1(e)). This behavior naturally points to the evaluation of the Fermi level and total energy (Figure 6 (d) (e)). The MP scheme demonstrates exceptional robustness in total energy calculations. Within a simple free-electron gas model, the kinetic energy contribution to the total energy is directly proportional to $\varepsilon_F$[49]. This relationship explains why the Fermi level shifts under MP smearing remain relatively small: the magnitude of the $\sigma$-induced variation in $\varepsilon_F$ across different smearing schemes scales directly with its perturbation to the total energy.

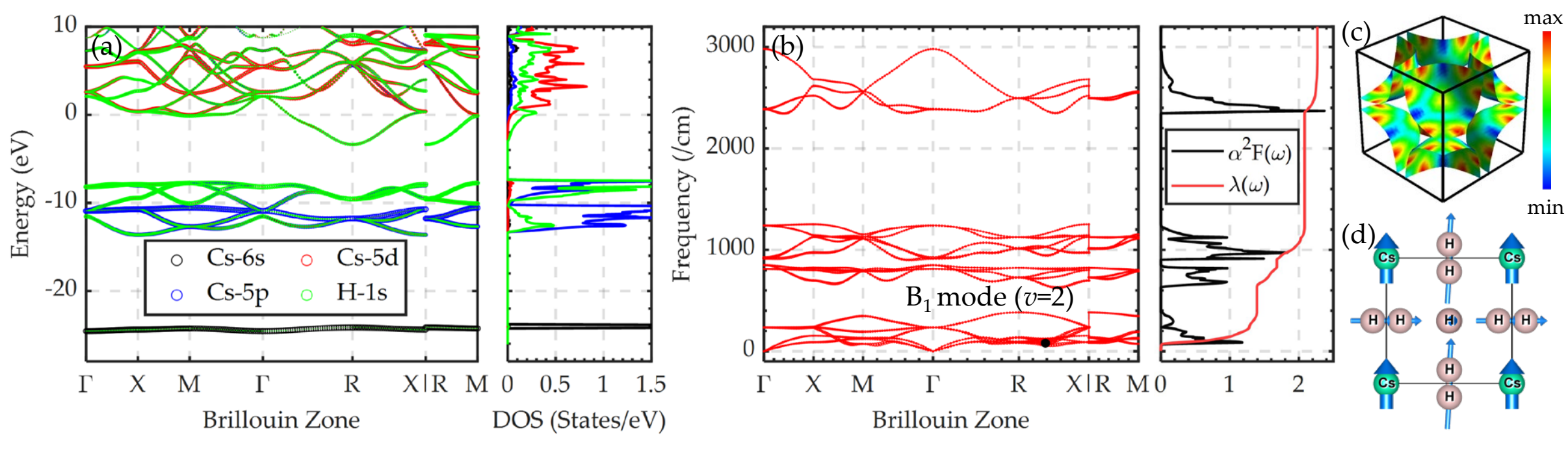

**Figure 7.** Superconducting mechanism of $CsH_6$ analyzed via the optimized tetrahedron method: (a) orbital-projected electronic band structure and corresponding density of states, (b) electron–phonon

coupling strength-projected phonon dispersion alongside the Eliashberg spectral function $\alpha^2F(\omega)$, (c) Fermi velocity-weighted Fermi surface, and (d) spatial visualization of the $B_1$ phonon mode.

Although smearing approximations introduce only mild systematic differences in $T_c$—and recently proposed DOS scaling approaches further improve $T_c$ robustness[24]—we aim to elucidate the impact of smearing schemes on the estimation of $\lambda$, as $\lambda$ provides direct insight into the fundamental EPC mechanism of conventional superconductors. Within the benchmark optimized tetrahedron method (Figure 7), the low-frequency phonons (0-400 $cm^{-1}$) of $CsH_6$ yield a dominant contribution of $\lambda = 1.39$ (61.2% of the total $\lambda$). In contrast, the intermediate-frequency (600-1300 $cm^{-1}$) and high-frequency (2300-3200 $cm^{-1}$) windows contribute only 0.69 (30.4%) and 0.19 (8.4%), respectively. Among the six low-frequency optical and acoustic modes, the three acoustic branches account for $\lambda_{\mathrm{Aco}} = 0.81$ (35.8% of the total), demonstrating that acoustic modes govern the EPC mechanism in $CsH_6$. Specifically, the second phonon mode $B_1$ near the midpoint of the R-X path exhibits the highest mode-resolved coupling strength $\lambda_{\mathbf{q}\nu}$. Spatial visualization reveals that this $B_1$ mode corresponds to concerted translations of $H_2$ molecules parallel or perpendicular to the Cs atoms, which directly modulates the ionic Cs-H bonding environment. Furthermore, valley-like electronic bands originating from H-1$s$ orbitals along the Γ--R--X path form pocket-like Fermi surfaces centered at the R point. In conclusion, the high-temperature superconductivity of 121 K in $CsH_6$ at ambient pressure originates from the strong coupling between H-1$s$ electrons and the soft $B_1$ phonon mode.

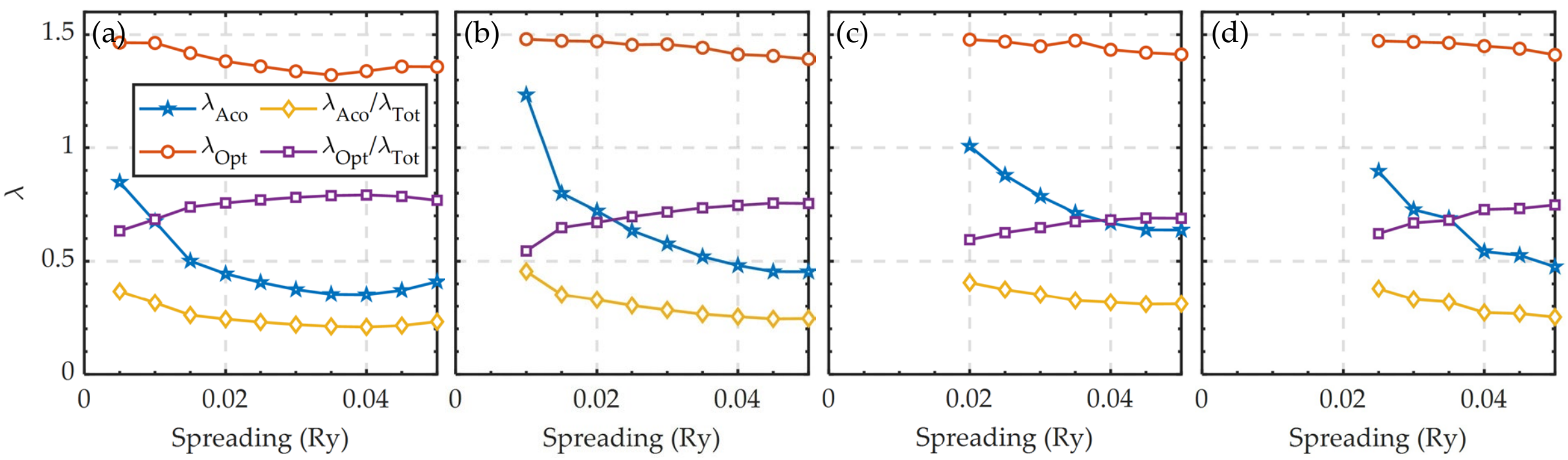

**Figure 8.** Smearing-width dependence of the electron–phonon coupling strength and its relative contributions, decomposed into acoustic and optical branches, for (a) Fermi–Dirac, (b) Gaussian, (c) Methfessel–Paxton, and (d) Marzari–Vanderbilt schemes.

Discrepancies among smearing approximations further manifest in the quantitative estimation of EPC from different parts. When decomposing the total coupling strength into acoustic $\lambda_{\mathrm{Aco}}$ and optical $\lambda_{\mathrm{Opt}}$ components, $\lambda_{\mathrm{Opt}}$ exhibits a remarkably flat response to variations in $\sigma$ (Figure 8). Across all four smearing schemes, $\lambda_{\mathrm{Opt}}$ remains anchored near 1.4, in excellent agreement with the benchmark value of $\lambda_{\text{Opt}} = 1.39$. This stability indicates that occupation smearing exerts a negligible effect on optical-branch EPC. In contrast, $\lambda_{\text{Aco}}$ displays a pronounced inverse dependence on $\sigma$. This behavior arises from two primary factors. First, the inverse-square dependence of coupling strength on phonon frequency ($\lambda \propto 1/\omega^2$) amplifies the sensitivity of low-frequency acoustic modes to $\sigma$, whereas high-frequency optical modes remain largely impervious. Second, while the optical modes stem entirely from H–H covalent bonding, the dominant $B_1$ mode governing the EPC mechanism is intimately tied to ionic Cs–H interactions. Although the cumulative optical contribution exceeds that of the acoustic branches, $\lambda_{\text{Opt}}$ represents a summation over 18 optical phonon branches, whereas $\lambda_{\text{Aco}}$ comprises only 3 acoustic branches. Consequently, the mode-averaged coupling weight of the acoustic branches is substantially higher, rendering $\lambda_{\text{Aco}}$ exceptionally sensitive to $\sigma$.

Additionally, we note the convergence behavior of $\lambda$. Employing a larger $\sigma$ requires a relatively sparse $k$-point mesh to achieve a converged $\lambda$, demonstrating that a long-tailed smearing function effectively accelerates $\lambda$ convergence. Once $\lambda$ reaches a convergence plateau, further increasing the $k$-point density no longer affects the calculated $\lambda$. However, the height of this convergence plateau systematically decreases as $\sigma$ increases, which aligns seamlessly with the monotonic downward trend of $\lambda$ with respect to $\sigma$ discussed in Figure 6(a). An excessively large $\sigma$ underestimates $\lambda$, whereas an overly small $\sigma$ overestimates $\lambda$ and may even induce dynamical instabilities. Consequently, a fundamental question emerges: how can we systematically determine the optimal smearing width $\sigma$ for different smearing functions to accurately match the benchmark $\lambda$ obtained from the optimized tetrahedron method?

Giving the free-electron gas model serves as an ideal framework for conceptualizing the electronic distribution in metals, the band dispersion at a wavevector $\mathbf{k}$ near the Fermi surface is expressed as:

$$\varepsilon(\mathbf{k}) = \frac{\hbar^2 k^2}{2m^*} \tag{16}$$

where $\hbar$ is the reduced Planck constant, $m^*$ is the electron effective mass, and $k = |\mathbf{k}|$ is the magnitude of the electronic wavevector. For a small variation in wavevector, $dk$, the derivative of the band energy with respect to $k$ is given by:

$$\nabla_{\mathbf{k}} \varepsilon = \frac{d\varepsilon}{dk} = \frac{\hbar^2 k}{m^*} = \hbar v_k \tag{17}$$

where $v_{\mathbf{k}}$ is the electronic group velocity. At the Fermi surface, we approximate $k \approx k_F$ and $v_{\mathbf{k}} \approx v_F$, where $k_F$ and $v_F$ represent the Fermi wavevector and Fermi velocity, respectively. For a given crystal lattice, the maximum $k$-point spacing $\Delta k_{\max}$ determined by the Monkhorst–Paxton grid generation scheme is expressed as[50]:

$$\Delta k_{\max} = \frac{2\pi}{(aN_k)_{\min}} \tag{18}$$

where $a$ denotes the lattice constant along a given crystallographic direction, and $N_k$ is the number of $k$-points sampled along that direction. The minimum value is obtained by computing $aN_k$ across all three spatial directions. Substituting this into the energy difference expression yields:

$$\Delta\varepsilon = \left|\nabla_{\mathbf{k}}\varepsilon \cdot \Delta\mathbf{k}\right| \tag{19}$$

which yields the maximum energy interval between adjacent $k$-points:

$$\Delta\varepsilon_{\max} = \left|\frac{d\varepsilon}{dk}\right|_{\max} \cdot \Delta k_{\max} = \hbar v_{F,\max} \cdot \frac{2\pi}{(aN_k)_{\min}} \tag{20}$$

Consequently, the smearing width employed must span the maximum energy interval between adjacent $k$-points to ensure that electronic states are fully smeared and continuously cover the Fermi surface. Specifically, the effective energy window of a given smearing function should match $\Delta\varepsilon_{\max}$; a width below $\Delta\varepsilon_{\max}$ leads to undersampling artifacts, whereas a width exceeding $\Delta\varepsilon_{\max}$ causes excessive dispersion of electronic occupations.

$$\beta\sigma = \Delta\varepsilon_{\max} = \frac{2\pi\hbar v_{F,\max}}{(aN_k)_{\min}} \tag{21}$$

where $\beta\sigma$ represents the effective energy window width, with numerical factors of $\beta = 7.33$, $2.77$, $1.49$, and $1.91$ determined for the FD, Gaussian, MP, and MV smearing function, respectively (Figure 1 (e)). Taking the widely used MP as a representative example, calibrating the calculated $\lambda$ for $CsH_6$ to match the benchmark tetrahedron method requires $\sigma = 0.03$ Ry (Figure 6 (a)). Substituting its $v_{F,\max} = 8.83 \times 10^5$ m/s, $(aN_k)_{\min} = 3.94\text{Å} \times 16$ and $\beta$=1.49 into Equation (21) yields a theoretical estimate of $\sigma = 0.028$ Ry, which agrees remarkably well with the optimal DFT setup. Recognizing that the single case of $CsH_6$ is insufficient to establish the general applicability of Equation (21), we performed external benchmark validations on three additional canonical systems: $MgB_2$, high-pressure fcc Li, and $PbNH_4B_6C_6$. For these three systems, the $\sigma$ values required in DFT calculations to reproduce the tetrahedron benchmark are $0.02$, $0.01$, and $0.03$ Ry, respectively, which align closely with the values calculated by Equation (21) with $0.019$, $0.014$, and $0.029$ Ry (Table S1).

In summary, benchmarked against the parameter-free optimized tetrahedron method, cubic $CsH_6$ exhibits a remarkable ambient-pressure $T_c$ of 121 K driven by strong H-1s electron coupling to the $B_1$ phonon mode. This excellent candidate, however, risks being identified as dynamically unstable and discarded in conventional high-throughput workflows when empirical smearing parameters induce acoustic imaginary frequency. We trace how numerical smearing propagates errors from electronic structure through phonon dynamics to electron-phonon coupling. Long-tailed smearing shifts the Fermi level and washes out van Hove singularities, whereas short-tailed smearing produces density-of-states oscillations that drive Fermi-surface-nesting instabilities via linear response. Because acoustic phonons dominate the EPC underlying this high-$T_c$ state, $\lambda_{Aco}$ fluctuates markedly while $\lambda_{Opt}$ remains robust. The $T_c$ estimate proves stable when $\varepsilon_F$ lies on the DOS slope or valley, yet demands finer $k$ meshes or larger $\sigma$ when pinned to the van Hove singularity. Finally, a free-electron-gas-derived $\sigma$ scaling relation, which equates the smearing window to the maximum energy spacing between adjacent $k$ points, provides an objective and computationally efficient criterion for high-throughput superconductors screening.

## Declaration of Competing Interest

The authors declare that they have no known competing financial interests or personal relationships that could have appeared to influence the work reported in this paper.

## Acknowledgements

R.-Q. Z. was supported by grants from the Research Grants Council of the Hong Kong SAR (CityU 11305618 and 11306219) and the National Natural Science Foundation of China (11874081). Y.-W. F. acknowledges the Extraordinary Grant of CSIC (Grant No. 2025ICT122) and the financial support received from the IKUR Strategy under the collaboration agreement between Ikerbasque Foundation and Centro de Física de Materiales (CFM-MPC) on behalf of the Department of Science, Universities and Innovation of the Basque Government (HPCAI21: AI-CrysPred). The authors also thank for the support of Open Source Supercomputing Center of S-A-I for providing the high-performance simulation.

## Supplementary materials

Supplementary materials to this article can be found online.

## Reference

[1] P. Raccuglia, K. C. Elbert, P. D. Adler, C. Falk, M. B. Wenny, A. Mollo, M. Zeller, S. A. Friedler, J. Schrier, and A. J. Norquist, Machine-learning-assisted materials discovery using failed experiments, Nature **533**, 73 (2016).

[2] Z. Wan and Q.-D. Wang, Machine learning prediction of the exfoliation energies of two-dimension materials via data-driven approach, The Journal of Physical Chemistry Letters **12**, 11470 (2021).

[3] Z. Wan, Q.-D. Wang, D. Liu, and J. Liang, Accelerating the optimization of enzyme-catalyzed synthesis conditions via machine learning and reactivity descriptors, Organic & Biomolecular Chemistry **19**, 6267 (2021).

[4] Z. Wan, Q. D. Wang, D. Liu, and J. Liang, Machine learning prediction of the optimal carrier concentration and band gap of quaternary thermoelectric materials via element feature descriptors, International Journal of Quantum Chemistry **121**, e26752 (2021).

[5] W. Kohn and L. J. Sham, Self-consistent equations including exchange and correlation effects, Physical review **140**, A1133 (1965).

[6] Z. Tong, S. Li, X. Ruan, and H. Bao, Comprehensive first-principles analysis of phonon thermal conductivity and electron-phonon coupling in different metals, Physical review B **100**, 144306 (2019).

[7] Y. Takimoto, F. Vila, and J. Rehr, Real-time time-dependent density functional theory approach for frequency-dependent nonlinear optical response in photonic molecules, The Journal of chemical physics **127** (2007).

[8] Z. Wan, W. Xu, T. Yang, and R. Zhang, As-Li electrides under high pressure: Superconductivity,

plastic, and superionic states, Physical Review B **106**, L060506 (2022).
[9] Z. Wan, G.-H. Zhong, R. Zhang, and H.-Q. Lin, The theory of “electride-carrier” precompression, Science China Physics, Mechanics & Astronomy **68**, 108262 (2025).
[10] Z. Wan, G.-H. Zhong, R. Zhang, and H.-Q. Lin, Renormalized quantum anharmonicity enhanced electron-phonon coupling in the ambient-pressure compound Rb H 6, Physical Review B **112**, L220504 (2025).
[11] W. Kohn, A. D. Becke, and R. G. Parr, Density functional theory of electronic structure, The journal of physical chemistry **100**, 12974 (1996).
[12] Z. Wan, Q.-D. Wang, D. Liu, and J. Liang, Effectively improving the accuracy of PBE functional in calculating the solid band gap via machine learning, Computational Materials Science **198**, 110699 (2021).
[13] P. Verma and D. G. Truhlar, Does DFT+ U mimic hybrid density functionals?, Theoretical Chemistry Accounts **135**, 182 (2016).
[14] M. Methfessel and A. T. Paxton, High-precision sampling for Brillouin-zone integration in metals, physical review B **40**, 3616 (1989).
[15] F. J. dos Santos and N. Marzari, Fermi energy determination for advanced smearing techniques, Physical Review B **107**, 195122 (2023).
[16] P. E. Blöchl, O. Jepsen, and O. K. Andersen, Improved tetrahedron method for Brillouin-zone integrations, Physical Review B **49**, 16223 (1994).
[17] M. Kawamura, Y. Gohda, and S. Tsuneyuki, Improved tetrahedron method for the Brillouin-zone integration applicable to response functions, Physical Review B **89**, 094515 (2014).
[18] J. J. Jorgensen and G. L. Hart, Effectiveness of smearing and tetrahedron methods: best practices in DFT codes, Modelling and Simulation in Materials Science and Engineering **29**, 065014 (2021).
[19] M. Kawamura, Y. Hizume, and T. Ozaki, Benchmark of density functional theory for superconductors in elemental materials, Physical Review B **101**, 134511 (2020).
[20] G. Kresse and J. Furthmüller, Efficiency of ab-initio total energy calculations for metals and semiconductors using a plane-wave basis set, Computational materials science **6**, 15 (1996).
[21] J. Arnaud, J. Boé, L. Chusseau, and F. Philippe, Illustration of the Fermi–Dirac statistics, American Journal of Physics **67**, 215 (1999).
[22] T. Björkman and O. Grånäs, Adaptive smearing for Brillouin zone integration, International Journal of Quantum Chemistry **111**, 1025 (2011).
[23] G. de Miranda Nascimento, F. J. dos Santos, M. Bercx, D. Grassano, G. Pizzi, and N. Marzari, Accurate and efficient protocols for high-throughput first-principles materials simulations, npj Computational Materials (2026).
[24] K. Bozier, K. Wang, B. Monserrat, and C. J. Pickard, High-throughput superconducting T c predictions through density of states rescaling, Physical Review B **113**, 064507 (2026).
[25] C. Pellegrini and A. Sanna, Ab initio methods for superconductivity, Nature Reviews Physics **6**, 509 (2024).
[26] I. A. Wrona, P. Niegodajew, and A. P. Durajski, High-Temperature Ternary Superhydrides: A Strategic Roadmap to Optimal Superconducting Parameters, Advanced Functional Materials **35**, 2423680 (2025).
[27] L. Boeri, R. Hennig, P. Hirschfeld, G. Profeta, A. Sanna, E. Zurek, W. E. Pickett, M. Amsler, R. Dias, and M. I. Eremets, The 2021 room-temperature superconductivity roadmap, Journal of Physics: Condensed Matter **34**, 183002 (2022).

[28] H. Tütüncü, G. Srivastava, S. Bağcı, and S. Duman, Theoretical examination of whether phonon dispersion in Nb 3 Sn is anomalous, Physical Review B—Condensed Matter and Materials Physics **74**, 212506 (2006).
[29] G. De Marzi, L. Morici, L. Muzzi, A. della Corte, and M. Buongiorno Nardelli, Strain sensitivity and superconducting properties of Nb3Sn from first principles calculations, Journal of Physics: Condensed Matter **25**, 135702 (2013).
[30] R. Zhang, P. Gao, X. Wang, and Y. Zhou, First-principles study on elastic and superconducting properties of Nb3Sn and Nb3Al under hydrostatic pressure, AIP Advances **5** (2015).
[31] R. Szczęśniak and A. P. Durajski, Unusual sulfur isotope effect and extremely high critical temperature in H3S superconductor, Scientific reports **8**, 6037 (2018).
[32] R. Markiewicz, Van hove singularities and high-T c superconductivity: a review, International Journal of Modern Physics B **5**, 2037 (1991).
[33] Y. Luo, Y. Han, J. Liu, H. Chen, Z. Huang, L. Huai, H. Li, B. Wang, J. Shen, and S. Ding, A unique van Hove singularity in kagome superconductor CsV3-x Ta x Sb5 with enhanced superconductivity, Nature communications **14**, 3819 (2023).
[34] S. Baroni, P. Giannozzi, and A. Testa, Green's-function approach to linear response in solids, Physical review letters **58**, 1861 (1987).
[35] S. Y. Savrasov and D. Y. Savrasov, Electron-phonon interactions and related physical properties of metals from linear-response theory, Physical Review B **54**, 16487 (1996).
[36] S. Baroni, S. de Gironcoli, A. Dal Corso, and P. Giannozzi, Phonons and related crystal properties from density-functional perturbation theory, Reviews of Modern Physics **73**, 515 (2001).
[37] P. Giannozzi, S. de Gironcoli, P. Pavone, and S. Baroni, Ab initio calculation of phonon dispersions in semiconductors, Physical Review B **43**, 7231 (1991).
[38] M. J. Gillan, Calculation of the vacancy formation energy in aluminium, Journal of Physics: Condensed Matter **1**, 689 (1989).
[39] M. Methfessel and A. T. Paxton, High-precision sampling for Brillouin-zone integration in metals, Physical Review B **40**, 3616 (1989).
[40] N. Marzari, D. Vanderbilt, A. De Vita, and M. C. Payne, Thermal Contraction and Disordering of the Al(110) Surface, Physical Review Letters **82**, 3296 (1999).
[41] M. Weinert and J. W. Davenport, Fractional occupations and density-functional energies and forces, Physical Review B **45**, 13709 (1992).
[42] G. Lehmann and M. Taut, On the Numerical Calculation of the Density of States and Related Properties, physica status solidi (b) **54**, 469 (1972).
[43] N. W. Ashcroft and N. D. Mermin, *Solid State Physics* (Holt, Rinehart and Winston, 1976).
[44] F. Giustino, Electron-phonon interactions from first principles, Reviews of Modern Physics **89**, 015003 (2017).
[45] G. Grüner, The dynamics of charge-density waves, Reviews of Modern Physics **60**, 1129 (1988).
[46] Y.-J. Feng, M.-J. Jiang, H.-B. Ding, H.-L. Tian, Z.-H. Lu, G.-H. Zhong, C.-L. Yang, X.-J. Chen, and H.-Q. Lin, High-Temperature Superconductivity in H3S up to 253 K at a Pressure of 140 GPa by Doping Holes, The Journal of Physical Chemistry C **126**, 20702 (2022).
[47] X. Wang, W. E. Pickett, M. Hutcheon, R. P. Prasankumar, and E. Zurek, Why Mg2IrH6 Is Predicted to Be a High-Temperature Superconductor, But Ca2IrH6 Is Not, Angewandte Chemie International Edition **63**, e202412687 (2024).
[48] P. B. Allen and R. C. Dynes, Transition temperature of strong-coupled superconductors reanalyzed,

Physical Review B **12**, 905 (1975).
[49] A. Sommerfeld, Zur Elektronentheorie der Metalle auf Grund der Fermischen Statistik, Zeitschrift für Physik **47**, 1 (1928).
[50] H. J. Monkhorst and J. D. Pack, Special points for Brillouin-zone integrations, Physical Review B **13**, 5188 (1976).